Book Chapter

# Filipino Students' Willingness to Use AI for Mental Health Support: A Path Analysis of Behavioral, Emotional, and Contextual Factors

John Paul P. Miranda[1*], Rhiziel P. Manalese[1], Ivan G. Liwanag[1], Rodel T. Alimurong[1], Alvin B. Roque[1]

1. **Pampanga State University**, Pampanga, Philippines

**\* Correspondence:**
John Paul P. Miranda, Pampanga State University, jppmiranda@pampangastateu.edu.ph

**How to cite this article:**
Miranda, J. P. P., Manalese, R. P., Liwanag, I. G., Alimurong, R. T., & Roque, A. B. (2026). Filipino Students' Willingness to Use AI for Mental Health Support: A Path Analysis of Behavioral, Emotional, and Contextual Factors. In A. ElSayary & A. Shomotova (Eds.), *Implications for Students' Mental Health in the Digital Age: AI and Cyber Behavior* (pp. 381-404). IGI Global Scientific Publishing.
https://doi.org/10.4018/979-8-3373-4222-1.ch015

**Article History:**
Submitted: 12 October 2025
Accepted: 04 February 2026
Published: 27 March 2026

## ABSTRACT

This study examined how behavioral, emotional, and contextual factors influence Filipino students' willingness to use artificial intelligence (AI) for mental health support. Results showed that habit had the strongest effect on willingness, followed by comfort, emotional benefit, facilitating conditions, and perceived usefulness. Students who used AI tools regularly felt more confident and open to relying on them for emotional support. Empathy, privacy, and accessibility also increased comfort and trust in AI systems. The findings highlight that emotional safety and routine use are essential in promoting willingness. The study recommends AI literacy programs, empathic design, and ethical policies that support responsible and culturally sensitive use of AI for student mental health care.

## INTRODUCTION

Artificial intelligence (AI) continues to shape new possibilities in mental health care. Studies show that it can identify early signs of psychological disorders such as depression, anxiety, schizophrenia, and bipolar disorder (Basha et al., 2025; Olawade et al., 2024). According to Olawade et al. (2024), AI can support professionals by analyzing behavioral data collected beyond clinical settings, providing deeper insight into an individual's everyday experiences and emotional states. Basha et al. (2025) also emphasized that AI technologies, including machine learning and deep learning, have been used to detect and diagnose mental health issues among students through behavioral, facial, and survey-based indicators (Baran & Cetin, 2025; Thakkar et al., 2024).

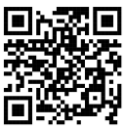

Beyond clinical use, AI applications in education have demonstrated their ability to personalize learning and improve access to information. Yuskovych-Zhukovska et al. (2022) described AI's potential to individualize instruction and analyze large-scale learning data to anticipate challenges and propose solutions. Zhou et al. (2022) further noted that AI systems can assist in tracking emotional and cognitive changes which could be helpful in providing psychological insights that benefit both learning and mental well-being.

Despite these technological opportunities, the Filipino context presents unique social and cultural challenges that influence mental health awareness and help-seeking behavior. Martinez et al. (2020) reported that Filipinos often show reluctance to acknowledge mental health concerns and prefer to seek support from family or friends instead of professionals. This reluctance reflects a deeply rooted value of self-reliance and a collective desire to avoid stigma. Yuduang et al. (2022) observed that many Filipinos remain unaware of existing digital mental health applications, even though these tools can manage mild to moderate symptoms effectively. Among students, Martinez et al. (2020) found that depressive symptoms are often normalized and attributed to ordinary life difficulties. Tan et al. (2025) added that students hesitate to seek professional help because of the stigma attached to mental health support, choosing self-management or peer guidance instead.

In light of these observations, this chapter aims to (1) describe the behavioral and emotional factors that influence Filipino students' willingness to use AI for mental-health support, (2) explain how comfort, habit, and emotional benefit relate to students' willingness, (3) discuss these relationships within the context of higher education in the Philippines, and (4) offer practical and policy recommendations that encourage the responsible and culturally aware use of AI-based mental-health tools.

## BACKGROUND

The growing usefulness of AI tools in education and mental health has been supported by several recent systematic reviews. After analyzing 85 studies, Cruz-Gonzalez et al. (2025) concluded that AI systems are effective in identifying, categorizing, and predicting mental health risks across different clinical contexts. Ni and Jia (2025) examined 36 empirical studies and reported that the use of AI in healthcare improves symptom tracking, patient engagement, and waiting time management. In another review of existing studies, Deckker and Sumanasekara (2025) found that AI contributes to the development of psychological engagement, self-awareness, empathy, and emotional regulation among students. Despite these positive findings, challenges remain in standardizing evaluation metrics and addressing algorithmic bias, cultural misinterpretations of emotions, and privacy issues. Studies have also noted that there is continued agreement that AI should be designed to complement, rather than replace, human therapeutic care (Sezgin, 2023; Spytska, 2025).

International research has also revealed continuing gaps between AI policy, implementation, and awareness within academic institutions. Studies from Australia, Saudi Arabia, India, and Europe conducted between 2022 and 2025 point to structural problems in educational systems. In a comprehensive analysis, Ullrich et al. (2022) identified four major concerns that limit the development of AI in education. These include the prioritization of administrative functions over instructional applications, the lack of interdisciplinary collaboration, the unequal representation of cross-national research, and the neglect of emerging areas of inquiry. The findings of other scholars support these conclusions. Todres and Sun (2025) found policy and practice inconsistencies based on interviews with Australian academics. Alharbi (2024) observed a mismatch between teachers' perceptions of AI use and students' actual engagement in Saudi

Arabia. In Europe, Titko et al. (2023) reported that more than half of academic staff recognized the importance of AI but lacked the technical skills to use it effectively.

In the Philippines, positive attitudes toward AI appear to be shaped by cultural values that emphasize emotional adaptability and collective harmony. Layugan et al. (2024) traced these attitudes to *Kapwa* Theory, which emphasizes shared identity and relational connection. They identified three emotional constructs that describe this orientation: *Madamdaming Pakikiangkop* (Emotional Adaptability), *Madamdaming Loob* (Emotional Identity), and *Madamdaming Pamamaraan* (Emotional Facilitation). The findings of Barnes et al. (2024) align with this perspective and indicate that collectivist cultures are more likely to view AI as a cooperative extension of the self rather than as a disruptive force. In local studies, de Leon et al. (2024) found that academic librarians welcomed AI as a tool for professional growth. Sibug et al. (2024) reported that teachers showed openness and readiness toward AI despite early hesitation. In contrast, Fabro et al. (2024) noted that students in academic writing contexts expressed neutral attitudes toward AI. These results suggest the presence of generational or situational differences in AI perception. Similar patterns of context-dependent acceptance and cautious openness toward AI tools among Filipino students have been documented in prior studies on AI-related learning in Philippine higher (Bringula et al., 2025; Hernandez et al., 2025; J. P. Miranda et al., 2024; Roga et al., 2025).

Students' routine cyber behaviors provide additional context for understanding these differences. Frequent engagement with digital platforms and AI-enabled tools shapes how students evaluate and adopt emerging technologies. Technology adoption models, particularly the Technology Acceptance Model and UTAUT/UTAUT2, explain these behaviors through performance expectancy, effort expectancy, social influence, facilitating conditions, and habit, which jointly predict intention and sustained use (Emon & Khan, 2025; Mustafa & Garcia, 2021; Sergeeva et al., 2025; Tbaishat et al., 2026). Empirical studies on AI adoption in higher education show that perceived usefulness and social influence are strong predictors of students' behavioral intentions, while habit explains continued engagement over time, especially in collectivist learning environments where peer endorsement and repeated exposure normalize technology use (Acosta-Enriquez et al., 2024; Aldreabi et al., 2025; Nurtanto et al., 2025; Valle et al., 2024). This perspective supports viewing Filipino students' willingness to use AI for mental-health support as an extension of established cyber practices shaped by social and cultural contexts.

Developing culturally grounded and locally relevant AI systems has become an important consideration in this context. Models that fail to recognize cultural nuances may lead to misinterpretations and ethical concerns when applied in real-world situations. This issue has been widely discussed in recent theoretical and review literature published. Ożegalska-Łukasik and Łukasik (2023) emphasized the importance of designing culturally sensitive AI systems for multicultural societies. Fuadi et al. (2025) discussed the limitations of Western-based emotion recognition models that overlook nonverbal cues common in other cultures. Barker et al. (2025) also pointed out that cultural bias is an ethical weakness that can be mitigated through region-specific adjustments and adaptive consent mechanisms. Bach et al. (2024) provided the strongest empirical support for these arguments through a review of studies that showed how user characteristics and cultural context influence trust in AI systems.

# MAIN FOCUS OF THE CHAPTER: BEHAVIORAL AND EMOTIONAL PATHWAYS INFLUENCING WILLINGNESS TO USE AI FOR MENTAL-HEALTH SUPPORT

## Methodology

A total of 536 Filipino college students who participated (Female = 364, Male = 168, non-binary = 4) reported using AI chatbot for emotional support/ stress-related purposes. All substantive items used a 5-point Likert scale (1 = Strongly disagree to 5 = Strongly agree). The constructs were: Perceived Usefulness (PU; 3 items), Ease of Use (EOU; 3), Social Influence (SI; 3), Facilitating Conditions (FC; 3), Emotional Benefit (EB; 3), Habit (HAB; 3), Empathic Response (ER; 3), Comfort (COM; 4), Willingness (WILL; 3), and Academic Stress (STRESS; 3).

The survey was administered online. Participation was voluntary and anonymous, and no personal identifying information was collected. Items were grouped by construct, and within-block order was randomized to reduce response patterns. Ethical norms for informed participation were observed.

Internal consistency was high for most multi-item constructs: ER α = .939, EB α = .930, EOU α = .915, HAB α = .898, PU α = .890, WILL α = .875, STRESS α = .876, FC α = .876, SI α = .858 COM α = 0.700. Bivariate correlations followed the expected pattern. Habit correlated most strongly with Willingness (approximately r ≈ .75), followed by Emotional Benefit (r ≈ .69) and Perceived Usefulness (r ≈ .62). Comfort and Facilitating Conditions also showed moderate positive relations to Willingness (both around r ≈ .54). These associations are consistent with a model in which behavioral routinization, perceived emotional gain, and perceived value anchor students' intention, while comfort and access provide enabling conditions.

To anticipate potential redundancy among predictors of Willingness, we computed Variance Inflation Factors (VIF) for the final predictor set (HAB, COM, EB, FC, PU). VIF values ranged roughly 1.5–2.8 and were well below conventional thresholds (VIF < 5), indicating no problematic multicollinearity in the planned path analysis. Following these checks, we estimated a path analysis with Willingness as the outcome. Direct paths were specified from HAB, COM, EB, FC, and PU to Willingness, in line with theory and the initial diagnostics. Internal pathways captured the emotional route (ER → EB → COM) and the behavioral route (PU, EOU, SI → HAB).

## First Issue: Emotional Resonance and Empathic Response

The ability of an AI system to show empathy has been recognized as a critical component in establishing effective human–computer interaction. In this study, empathic response refers to how students perceive the AI's capacity to understand their feelings, recognize emotional cues, and respond with sensitivity. When empathy is communicated through digital conversation, users experience a sense of connection that transforms a neutral interface into a perceived companion.

Prior work on social presence, the "computers as social actors" perspective, and affective computing suggests that perceived empathy increases users' emotional payoff and willingness to disclose. In mental-health contexts, empathic language and supportive reflections are associated with perceived relief and comfort, which in turn predict Willingness to seek support. Guided by this stream, we hypothesized Empathic Response → Emotional Benefit (H1a), Emotional Benefit → Comfort (H1b), and Comfort → Willingness (H1c).

As shown in Figure 1, Empathic Response → Emotional Benefit (β = .35, p < .001), Emotional Benefit → Comfort (β = .19, p = .001), and Comfort → Willingness (β = .18, p < .001). These links indicate that empathy-driven emotional relief enhances comfort, which in turn increases willingness to use AI for mental-health support.

These findings support prior studies that emphasize the importance of emotional intelligence in educational and therapeutic AI systems. The results demonstrate that empathy functions as an initial gateway to emotional engagement. Once students feel emotionally understood, they are more inclined to continue interacting with the system, thereby reinforcing comfort and willingness.

**Figure 1.**
*Pathway of Emotional Resonance from Empathic Response to Comfort and Willingness*

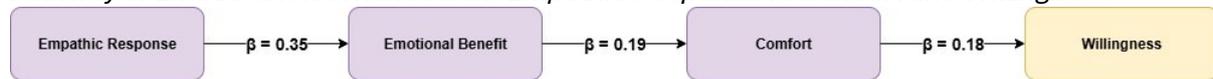

The observed pattern resonates with Filipino cultural orientations. Because *hiya* (a sense of modesty or shyness) and fear of social judgment often discourage open discussion of mental distress, an empathic AI may serve as a non-threatening intermediary. Students perceive the AI as a "listener" that does not judge or stigmatize, allowing them to disclose feelings they might otherwise suppress. This suggests that empathy-driven design is not merely a technical feature but a culturally relevant mechanism that increases users' comfort in seeking digital mental-health support.

## Second Issue: Habit as a Bridge to Willingness

A person's willingness to adopt a technology often depends on the degree to which its use becomes habitual. In this study, habit refers to repeated and effortless engagement with AI tools that leads to routine behavior. Once a behavior becomes automatic, it requires less deliberate motivation and is more likely to be sustained.

Models of technology acceptance (e.g., TAM/UTAUT2) and the psychology of automaticity posit that perceived usefulness, social endorsement, and low effort facilitate repeated use, which consolidates into habit; habit then becomes a proximal driver of Willingness. Accordingly, we hypothesized Perceived Usefulness → Habit (H2a), Social Influence → Habit (H2b), Ease of Use → Habit (H2c), and Habit → Willingness (H2d).

Figure 2 shows Perceived Usefulness → Habit ($\beta$ = .39, $p$ < .001), Social Influence → Habit ($\beta$ = .29, $p$ < .001), Ease of Use → Habit ($\beta$ = .14, $p$ = .002), and Habit → Willingness ($\beta$ = .41, $p$ < .001). These results confirm habit as the behavioral bridge between cognitive evaluations and willingness.

**Figure 2.**
*Behavioral pathway from Perceived Usefulness, Ease of Use, and Social Influence to habit and Willingness*

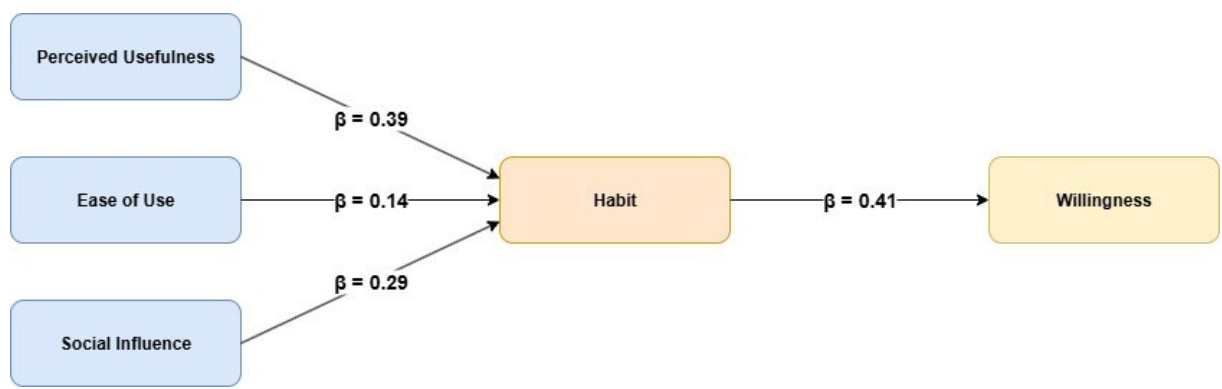

The results highlight that habit functions as a behavioral anchor for continued engagement. Filipino students who have previously used AI chatbots for academic purposes (e.g., study assistance or productivity tasks) may develop a sense of familiarity that transfers to emotional or mental-health use. This behavioral spillover suggests that comfort with AI technology in one domain may facilitate its acceptance in another.

From a cultural perspective, habit is particularly relevant in the Philippine setting, where consistency and community endorsement strengthen behavioral commitment. Students often rely on social cues (e.g., classmates' experiences or online testimonials) when deciding whether to try new tools. For this reason, that fostering positive and repeated exposure to AI mental-health systems can reinforce usage habits and enhance overall willingness.

### Third Issue: Comfort, Stigma, and Digital Disclosure

Comfort refers to the level of ease, emotional safety, and non-judgmental space students feel when engaging with AI about their mental state. When users feel comfortable, they are more likely to express emotions openly and explore the system's features without fear of stigma or embarrassment.

Research on self-disclosure, privacy calculus, and the online disinhibition effect indicates that perceived empathy and experienced relief reduce social threat and increase comfort in sharing sensitive information. In collectivist settings where stigma is salient, comfort functions as a key mediator translating supportive exchanges into help-seeking Willingness. Based on this logic, we hypothesized Empathic Response → Comfort (H3a), Emotional Benefit → Comfort (H3b), and Comfort → Willingness (H3c).

As depicted in Figure 3, Empathic Response → Comfort ($\beta = .47$, $p < .001$) and Emotional Benefit → Comfort ($\beta = .19$, $p = .001$), with Comfort → Willingness ($\beta = .18$, $p < .001$). Comfort therefore functions as the emotional bridge that transforms supportive interactions into willingness.

**Figure 3.**
*Emotional Pathway from Empathic Response and Emotional Benefit to Comfort and Willingness*

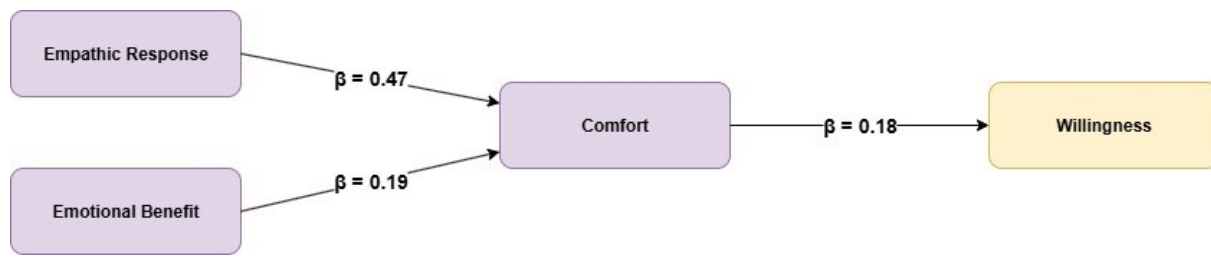

These findings indicate that comfort is the emotional bridge that transforms empathy and satisfaction into Willingness. When an AI system demonstrates understanding and responsiveness, it helps create a sense of privacy and acceptance that traditional social interactions may not provide. This dynamic is particularly relevant in the Filipino cultural context, where *hiya* (modesty) and *takot sa panghuhusga* (fear of being judged) often discourage students from seeking emotional help. The AI system, being nonhuman, reduces these social barriers which makes emotional disclosure more approachable and less intimidating.

The results further suggest that design strategies for AI mental-health tools must prioritize comfort-oriented interactions. This includes the use of warm, encouraging language, the avoidance of overly clinical phrasing, and sensitivity to the user's mood. By doing so, developers can strengthen the trust and emotional safety needed for sustained engagement.

### Fourth Issue: Access and Readiness

Access and readiness determine whether students can fully benefit from the emotional and behavioral advantages of AI-based mental-health systems. Here, access refers to device availability, stable connectivity, and opportunities to explore AI tools, while readiness pertains to confidence and familiarity in using these technologies; in this chapter, readiness is used descriptively rather than as a modeled mediator.

The facilitating-conditions construct in acceptance models and digital-divide findings show that resource availability and perceived value can influence adoption directly, especially for essential services like health support. When infrastructure is adequate and the tool is viewed as beneficial, Willingness increases without requiring an intermediate readiness mechanism. Thus, we hypothesized Facilitating Conditions → Willingness (H4a) and Perceived Usefulness → Willingness (H4b).

Figure 4 shows Facilitating Conditions → Willingness (β = .11, p = .003) and Perceived Usefulness → Willingness (β = .09, p = .028) as direct effects on willingness. No Readiness → Willingness path was estimated.

**Figure 4.**
*Direct effects of facilitating conditions and perceived usefulness on willingness to use AI for mental-health support*

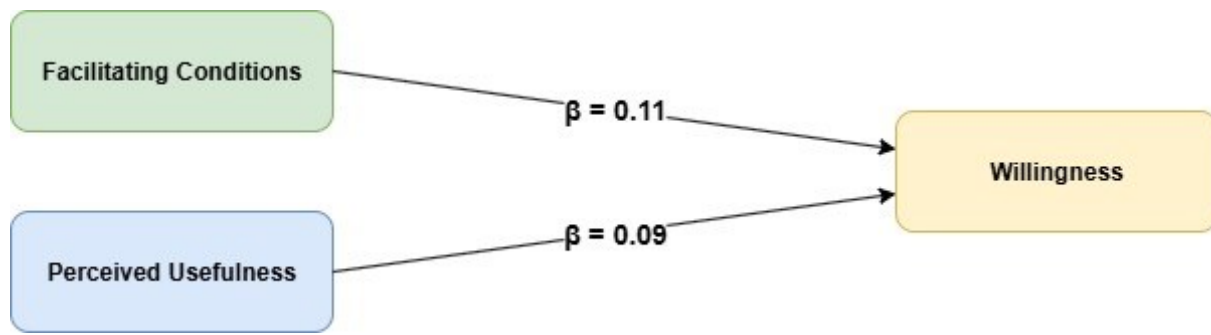

These findings emphasize that structural readiness (i.e., reliable technology, connectivity, and basic support) remains an essential prerequisite for emotional and behavioral engagement. Students may perceive AI systems as helpful or empathic, but without consistent access, these benefits cannot translate into habitual or emotionally sustained use. Psychological readiness also matters: students are more willing to engage when they feel in control of the interaction and trust the system to protect their privacy.

In the Philippine context, accessibility and readiness are influenced by socioeconomic disparities across regions and institutions. Students in urban universities may have more consistent exposure to AI tools than those in rural areas. Despite these differences, the results indicate that when students are provided with reliable access and guided exposure, their openness to AI-assisted mental-health support increases.

Hence, institutional readiness should accompany technological access. Universities should complement infrastructure with orientations or seminars on AI literacy and mental-health awareness. These measures improve usability, strengthen confidence, and support responsible engagement with AI-based mental-health tools.

### Path Model Summary

The final model explains 65.7% of the variance in Willingness ($R^2$ = 0.657; Table 1). Direct predictors are Habit (β = 0.41, p < .001), Comfort (β = 0.18, p < .001), Emotional Benefit (β = 0.11, p = .024), Facilitating Conditions (β = 0.11, p = .003), and Perceived Usefulness (β = 0.09, p = .028).

**Table 1.**
*Summary of Direct Predictors of Willingness to Use AI for Mental-Health Support*

| Path | β | p-value | Interpretation |
| --- | --- | --- | --- |
| Habit → Willingness | 0.41 | < .001 | Strongest positive effect |

**Miranda et al. (2026)**

| Path | β | p-value | Interpretation |
|---|---|---|---|
| Comfort → Willingness | 0.18 | < .001 | Significant positive effect |
| Emotional Benefit → Willingness | 0.11 | 0.024 | Significant positive effect |
| Facilitating Conditions → Willingness | 0.11 | 0.003 | Significant positive effect |
| Perceived Usefulness → Willingness | 0.09 | 0.028 | Significant positive effect |
| Model $R^2$ | 0.657 | — | High explanatory power |

As shown in Figure 5, Willingness is shaped by mutually reinforcing behavioral, emotional, and contextual influences. On the behavioral pathway, Perceived Usefulness, Ease of Use, and Social Influence promote Habit, which exerts the largest direct effect on Willingness. On the emotional pathway, Empathic Response increases Emotional Benefit, which enhances Comfort; comfort subsequently raises Willingness. On the contextual pathway, Facilitating Conditions directly support Willingness by ensuring dependable access and basic support. In addition to its indirect role via habit, Perceived Usefulness also has a small direct effect on Willingness (β = .09).

Functional roles of the predictors. Habit serves as the proximal driver of Willingness. By consolidating earlier evaluations into automatic, low-effort engagement, habit yields the largest direct effect (i.e., once students routinely consult an AI tool during stressful moments, sustained use relies less on persuasion and more on established behavior). Comfort functions as the model's emotional safety gate: beyond perceived usefulness, students must feel safe and unjudged to choose AI for sensitive concerns, which explains its direct link to Willingness. Emotional Benefit operates as an affective catalyst; immediate relief and reassurance during interaction motivate students to return, both directly and by strengthening Comfort. Facilitating Conditions provide the structural enabler; reliable access, connectivity, and basic support keep all routes operational and also exert an independent direct effect on Willingness. Perceived Usefulness follows a dual route: it provides a modest direct evaluative push to Willingness while exerting a larger indirect influence by seeding Habit.

Internal paths supporting these roles. The Empathic Response → Emotional Benefit link acts as the emotional ignition of the model; sensitive, attuned AI replies yield emotional relief, which then converts into Comfort (Emotional Benefit → Comfort). On the behavioral side, Perceived Usefulness, Social Influence, and Ease of Use provide the scaffolding of routine (i.e., value, normative endorsement, and low effort together promote Habit), which in turn drives Willingness.

**Figure 5.**
*Integrated path model of behavioral, emotional, and contextual predictors of willingness to use AI for mental-health support*

**Miranda et al. (2026)**

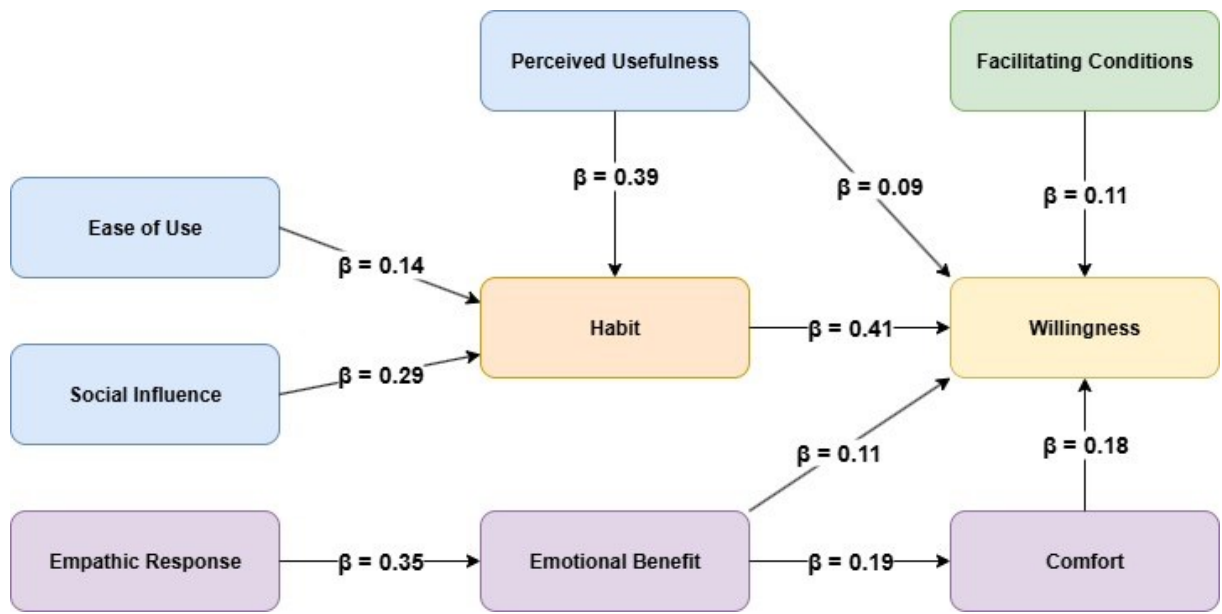

While the preceding model highlights the behavioral, emotional, and contextual drivers of willingness, the use of AI chatbots for mental health support also raises ethical and practical concerns that warrant consideration. Research on conversational AI in mental health has documented a range of ethical and safety issues, including privacy, data security, algorithmic bias, and the need for transparent governance, particularly when sensitive emotional information is disclosed (Bagane et al., 2026; Garcia, 2023; Iftikhar et al., 2025; Rahsepar Meadi et al., 2025; WHO, 2024). Studies of user perspectives further report concerns related to AI's limited empathic capacity, technical constraints, and risks of overreliance or dependency, especially when chatbots are perceived as substitutes for human support rather than as supplementary tools (Chan, 2025; Lee et al., 2025; Wang et al., 2025). Additional risks include misinterpretation or oversimplification of emotional states, as AI systems lack clinical judgment and may inadequately respond to complex or severe psychological distress which may reinforce the importance of trust and ethical safeguards in AI-mediated mental health contexts (Chan, 2025; Fiske et al., 2019; Garcia, 2024; Head, 2025; Lee et al., 2025). Empirical work on AI health chatbots also identifies concrete security and privacy vulnerabilities that can undermine user confidence and raise compliance concerns (Li, 2023; Yener et al., 2025). Within students' broader digital help-seeking behaviors, AI chatbots function most appropriately as low-threshold extensions of online self-help practices rather than as replacements for professional care. This positioning reflects documented emotional strain and stress experiences among Filipino university students, which often motivate the use of accessible and private forms of support (De Nieva et al., 2021; Miranda & Tolentino, 2023; Serrano & Reyes, 2022), and highlights the need for ethical safeguards, clearly defined system boundaries, and structured referral mechanisms that direct users to qualified mental health professionals when distress exceeds the intended scope of AI-based support (AlMaskari et al., 2025; Aziz et al., 2025; Chin et al., 2023; Cruz-Gonzalez et al., 2025; Giray et al., 2024; Putica et al., 2025; Saeidnia et al., 2024).

## SOLUTIONS AND RECOMMENDATIONS

Higher education institutions can integrate AI chatbots for student support in a responsible and evidence-based manner. Universities should position these tools as supplements to existing counseling and wellness services. AI chatbots can be embedded in learning management systems or institutional portals to help monitor student well-being, provide stress management advice, and guide students toward self-help resources. To ensure ethical use, universities must implement AI literacy and digital well-being programs for both faculty and students. These programs should explain how AI systems collect and use data, the limits of their emotional understanding, and the importance of maintaining confidentiality. Institutions should also establish clear data management policies, define referral procedures to professional counselors, and form ethics committees to oversee AI-assisted support.

Developers and designers should apply human-centered principles that prioritize empathy, accessibility, and cultural sensitivity. AI-based mental health tools must demonstrate the capacity to recognize and respond to users' emotions through appropriate language, tone, and timing. Developers should include bilingual interaction features in both English and Filipino to reflect how students naturally communicate their emotions. Data security must remain a central feature of design, and users should be informed about how their information is stored and protected. Local cultural elements can also strengthen emotional resonance. Incorporating Filipino values such as respect, modesty, and shared empathy can help users feel comfortable and understood. Collaboration with psychologists, educators, and cultural experts is important to ensure that system dialogue and tone align with Filipino social and emotional norms.

Policymakers and governing agencies such as CHED, DOH, and DICT should create coherent guidelines for integrating AI into mental health initiatives while safeguarding student privacy. CHED can include AI-enabled mental health support in university digital transformation policies. DOH should provide professional and ethical standards for AI-assisted mental health applications. DICT must ensure that cybersecurity and data privacy measures are consistent with national laws. Together, these agencies can develop a national framework that regulates AI systems, promotes fairness, and protects users from harm. Regular evaluations, transparency in algorithms, and informed consent procedures can help ensure that AI systems remain ethical and effective in serving students.

Researchers and practitioners can use the findings of this study to guide future interdisciplinary efforts that combine counseling expertise and technological innovation. The results highlight the importance of empathy, habit, comfort, and access as predictors of students' willingness to use AI for mental health support. Researchers can explore how these factors influence long-term emotional well-being, while practitioners can integrate AI literacy and ethical awareness into counselor education programs. Collaboration among educators, psychologists, and data scientists can help create adaptive systems that balance human care with technological support. By combining academic evidence, ethical practice, and cultural understanding, AI systems can become effective tools for promoting emotional wellness among Filipino students.

## FUTURE RESEARCH DIRECTIONS

Future studies may further explore how continuous interaction with AI-based mental-health systems affects students' emotional well-being and coping behaviors. Longitudinal research designs are recommended to observe changes in emotional comfort and habit over time. Experimental approaches may also validate the mediating roles of empathy, habit, and comfort in predicting willingness. Likewise, mixed-method studies may uncover deeper insights into students' emotional experiences and patterns of digital disclosure. Comparative investigations between public and private higher-education institutions, or across rural and urban settings, may reveal contextual differences in access and readiness. Finally, interdisciplinary research involving educators, psychologists, and computer scientists is encouraged to refine AI systems that are ethical, culturally grounded, and emotionally responsive to Filipino learners.

## CONCLUSION

This chapter presented the behavioral, emotional, and contextual pathways that influence Filipino students' willingness to use AI for mental-health support. The results showed that habit, comfort, and emotional benefit are significant predictors of willingness, supported by empathic interaction and sufficient facilitating conditions. These findings indicate that emotional resonance complements technological usefulness in sustaining students' engagement with AI systems. When empathy and accessibility are integrated into AI design, students perceive such systems as supportive companions rather than impersonal tools. Hence, universities, policymakers, and developers must ensure that the use of AI for mental-health support remains ethical, inclusive, and culturally appropriate. A well-designed and contextually informed AI system can serve as a valuable complement to traditional counseling which helps in promoting emotional well-being and responsible technology use among Filipino students.

**Miranda et al. (2026)**

# KEY TERMS AND DEFINITIONS

**AI Writing Tools**: Computer-based tools and systems, such as ChatGPT, QuillBot, and Grammarly, used to assist academic writing through functions like text generation, grammar correction, summarization, and citation support.

**Miranda et al. (2026)**

**Artificial Intelligence** AI: A computer system that performs human-like reasoning and provides adaptive support for users.

**Willingness**: The readiness of students to use AI tools for emotional or mental health support.

**Habit**: A repeated behavior that leads to automatic and consistent use of AI systems.

**Comfort**: A state of emotional safety and ease when interacting with AI about mental health concerns.

**Emotional Benefit**: The positive emotional relief or reassurance gained from using AI for support.

**Empathic Response**: The ability of an AI system to understand and respond to the user's emotions.

**Perceived Usefulness**: The belief that AI systems are helpful in improving emotional well-being.

**Facilitating Conditions**: The resources and support that allow students to access and use AI tools.

**Ease of Use**: The simplicity and convenience of operating an AI system.

**Social Influence**: The effect of peers and social groups on students' decisions to use AI for support.

**Behavioral Pathway**: The route showing how usefulness and social influence form habit and increase willingness.

**Emotional Pathway**: The route showing how empathy and emotional benefit enhance comfort and willingness.

**Contextual Pathway**: The route showing how access, readiness, and support influence AI adoption.

**Filipino Cultural Sensitivity**: The consideration of Filipino values and emotions, such as *hiya* and *pakikiramdam*, in AI design and communication.